\documentclass[runningheads]{llncs}

\makeatletter
\newcommand{\comp}{\hspace{0pt}\nolinebreak-\hspace{0pt}}
\newsavebox{\@brx}
\newcommand{\llangle}[1][]{\savebox{\@brx}{\(\m@th{#1\langle}\)}%
  \mathopen{\copy\@brx\kern-0.5\wd\@brx\usebox{\@brx}}}
\newcommand{\rrangle}[1][]{\savebox{\@brx}{\(\m@th{#1\rangle}\)}%
  \mathclose{\copy\@brx\kern-0.5\wd\@brx\usebox{\@brx}}}

\usepackage[T1]{fontenc}
\usepackage{graphicx}
\usepackage{hyperref}

\usepackage{amsmath}
\usepackage{amssymb}
\usepackage{paralist}
\usepackage{multirow}
\usepackage{enumitem}
\usepackage{placeins}
\usepackage{xcolor}
\usepackage{tikz}
\usetikzlibrary{automata, positioning}

\usepackage{pgfplots}
\pgfplotsset{compat=1.18}
\usepackage{pgfplotstable}
\usepackage{filecontents}
\usepackage{wrapfig}

\usepackage{algorithm}
\usepackage{algpseudocode}
\usepackage{algorithmicx}
\usepackage{listings}
\usepackage[utf8]{inputenc}
\usepackage[small]{caption}
\begin{document}
\title{A Model Checker for Natural Strategic Ability}
%
%
\author{Marco Aruta\inst{1}
\and
Vadim Malvone\inst{2}
\and
Aniello Murano\inst{1}
}
\authorrunning{M. Aruta et al.}
%
\institute{Università degli Studi di Napoli Federico II, Italy \and
Télécom Paris, Institut Polytechnique de Paris, France}
%
\maketitle              
\begin{abstract}
In the last two decades, \textit{Alternating-time Temporal Logic} (\textit{ATL}) has been proved to be very useful in  modeling strategic reasoning for \textit{Multi-Agent Systems} (\textit{MAS}). However, this logic struggles to capture the bounded rationality inherent in human decision-making processes. To overcome these limitations, \textit{Natural Alternating-time Temporal Logic} (\textit{NatATL}) has been recently introduced. As an extension of \textit{ATL}, \textit{NatATL} incorporates bounded memory constraints into agents' strategies, which allows to resemble human cognitive limitations. 
In this paper, we present a model checker tool for \textit{NatATL} specifications - both for memoryless strategies and strategies with recall - integrated into \textit{VITAMIN}, an open-source model checker designed specifically for \textit{MAS} verification. By embedding \textit{NatATL} into \textit{VITAMIN}, we transform theoretical advancements into a practical verification framework, enabling comprehensive analysis and validation of strategic reasoning in complex multi-agent environments. Our novel tool paves the way for applications in areas such as explainable AI and human-in-the-loop systems, highlighting \textit{NatATL}'s substantial potential. 

\end{abstract}
\section{Introduction}
\textit{Multi-Agent Systems (MAS)} have gained significant attention in recent years due to their ability to model and analyze complex systems composed of multiple interacting agents~\cite{wooldridge2009introduction,jamroga2015logical}. These agents can be either human or artificial, and their interactions can range from cooperative to adversarial. Strategic reasoning plays a crucial role in understanding and predicting agent behavior and in designing systems that exhibit desired properties~\cite{MogaveroMPV14}. 

Over the past twenty years, formal verification works on \textit{MAS} have flourished, with practical applications developed by both theorists and practitioners \cite{lomuscio2017mcmas}. One of the most successful logics for representing agent conduct is \textit{Alternating-time Temporal Logic} (\textit{ATL}) \cite{ATLpaper}, which uses a strategic operator $\langle\!\langle A\rangle\!\rangle \phi$ to indicate that a coalition of agents $A$ has a strategy capable of enforcing $\phi$ regardless of the actions of all other agents. 

Recently, \textit{NatATL} ~\cite{natatl2017} has been introduced as a logic to better capture how the humans naturally strategize. \textit{NatATL} is a variant of ATL that updates the strategic operator $\langle\!\langle A\rangle\!\rangle \phi$ with a bounded version $\langle\!\langle A\rangle\!\rangle^{\leq k} \phi$, where $k \in \mathbb{N}$ denotes the complexity bound.
\textit{NatATL} has the potential to be useful in various AI applications, such as human-in-the-loop systems and explainable AI. Despite the potential of \textit{NatATL}, it has never been implemented in a tool. 

\paragraph{Our Contribution.}
This paper introduces, for the first time, a verification tool for \textit{NatATL}, both in the context of agents' strategies with or without recall. Our tool extends \textit{VITAMIN}~\cite{vitamin}, an open-source model checker recently released and tailored to the verification of \textit{MAS}. The implementation of \textit{NatATL} model checking we provide includes the following features:
\begin{inparaenum}[(i)]
    \item \textit{Strategies Generation}: The first step involves generating all strategies for each coalition whose complexity does not exceed a fixed value;
\item \textit{Model Pruning}: This phase involves projecting the strategies generated in the previous phase to simplify the model;
\item \textit{Model Checking CTL}: Since the pruning phase has made the problem independent from the coalition due to the projection of the strategy coalition, it is now possible to use CTL model checking. As the complexity bound on the size of natural strategies is constant, the problem is polynomial in the size of the model ~\cite{clarke1981design}.
\end{inparaenum}
Furthermore, this approach allows the synthesis of optimal strategies that lead to a solution. To enhance performance, a real-life usage scenario is presented, where this module is integrated into a more general framework that also incorporates an ATL verification module~\cite{Giulia2023}. These features facilitate the analysis and validation of complex multi-agent environments, paving the way for applications in many AI areas, 
demonstrating substantial potential for real-world impact.

\paragraph{Outline.}
The rest of the paper is organized as follows. Section~2 recalls the main notions of natural strategies and NatATL. Section~3 presents the implementation of our tool.
Section~4 evaluates the performance of our module by means of experiments. Finally, we conclude in Section~5.

\paragraph{Related Works.} 
Since model checking was first introduced, tools for formal verification have been essential in addressing the complexities of strategic reasoning in \textit{MAS}. Several notable tools have been developed to support this analysis. One of the foundational tools is \textit{NuSMV} \cite{cimatti1999nusmv}, a symbolic model checker for temporal logics such as \textit{CTL} and \textit{LTL}. It provides a framework for verifying finite state systems and has been widely used in both academic and industrial settings. Alongside \textit{NuSMV}, other model checkers like \textit{SPIN} \cite{holzmann1997SPIN}, which focuses on verifying distributed software systems, and \textit{UPPAAL} \cite{behrmann2006uppaal}, designed for real-time systems, have significantly contributed to the field of formal verification. In the realm of reactive and open-source systems, \textit{MOCHA} stands out. \textit{MOCHA} is a model checking tool for concurrent systems, enabling the specification and verification of system properties expressed in \textit{ATL}. \textit{MOCHA} was particularly innovative in its ability to explicitly handle strategy depth, making it one of the few tools capable of exploring the nuances of strategic reasoning \cite{mocha}. For \textit{MAS}, \textit{MCMAS} 
 \cite{MCMAS} is a widely recognized tool. Initially conceived for \textit{ATL}, \textit{MCMAS} has been extended to support various logics and verification tasks. Extensions include \textit{SLK}, which adds support for epistemic logic, \textit{TeslaMCMASX} for temporal epistemic strategic logics, \textit{MCMAS-Dynamic} for dynamic multi-agent systems, and \textit{MCMAS-T} for temporal logic with past operators. These extensions have broadened the applicability of \textit{MCMAS}, making it a versatile tool for analyzing strategic interactions in MAS \cite{SLK,TeslaMCMASX,MCMAST,MCMASDynamic}. Thus, as recent additional verification tools, we recall \textit{EVE} \cite{eve} and \textit{STV} \cite{KurpiewskiJK19}. Despite the advancements brought by these tools, most do not address the bounded rationality of agents, an essential aspect of real-world decision-making. This gap is where \textit{MOCHA}’s exploration of strategy depth was pioneering, but it did not fully capture bounded rationality. The need for tools that can synthesize strategies considering bounded rationality led to the development of \textit{VITAMIN} \cite{vitamin}, a user-friendly, open-source tool designed to overcome limitations of previous model checkers. \textit{VITAMIN} is extendable to various types of models and logic, making it suitable for \textit{NatATL}. \textit{VITAMIN}’s comprehensive approach to verification in \textit{MAS}, coupled with its flexibility, positions it as a critical tool for the formal analysis of strategic \textit{reasoning}. In this work, we build our module on VITAMIN, leveraging its strengths to extend its capabilities for \textit{NatATL}. By doing so, we aim to address the gaps left by previous tools and provide a robust solution for synthesizing strategies in \textit{MAS}, considering the bounded rationality of agents. This extension not only enhances the practical applicability of \textit{VITAMIN} but also contributes to the broader field of strategic reasoning in \textit{MAS}.

\section{Preliminaries}
\paragraph{VITAMIN tool.}
It is constructed for easy extension to accommodate various logics (for specifying the properties to verify) and models (for determining on what to verify such properties). This tool is implemented in Python - the same programmed language used for its modules creation - and its Graphical User Interface is accessible through a web browser \href{https://vitamin-app.streamlit.app/}{(https://vitamin-app.streamlit.app/)}, which makes the tool cross-platform. This accessibility is facilitated by utilising the Streamlit Python library, which supports the transparent sharing of Python programs via
HTTP protocol. This model checker supports the interaction with both expert and non-expert users.

\paragraph{NatATL.} Natural Alternating\comp Time Temporal Logic (\textit{NatATL})\cite{natatl2019} is a logic for natural strategic ability that enhances \textit{ATL} by integrating human reasoning constraints, thereby expanding its applicability and effectiveness. \textit{NatATL} effectively addresses usability issues related to the functional requirements of reactive systems. 
It considers situations where multiple agents, each with their own goals and capabilities, can take actions concurrently.
\textit{NatATL} syntax derives from substituting the modality $\langle\!\langle A \rangle\!\rangle$ in \textit{ATL} with the bounded strategic modality $\langle\!\langle A \rangle\!\rangle^{\leq k}$. Intuitively, $\langle\!\langle A \rangle\!\rangle^{\leq k} \gamma$ says that a coalition of agents ($A \!\subseteq \!Agt$) has a collective natural strategy of size at most $k$ to enforce property $\gamma$. Similar to \textit{ATL}, \textit{NatATL} formulas predicate over a set of atomic propositions $AP$ and uses classical temporal operators. Therefore, the \textit{NatATL} language can be delineated by the following grammar: $\varphi ::= p \,|\, \neg\varphi \,|\, \varphi \land \varphi \,|\, \langle\!\langle A \rangle\!\rangle^{\leq k} X \varphi \,|\, \langle\!\langle A \rangle\!\rangle^{\leq k}\, \varphi \, U \varphi$; where $p$ is an atomic proposition, $k$ is the complexity bound, $\varphi$ is a composed formula and, X and U are temporal operators and stand for "next" and "until", respectively. 
For a coalition $A$, $s_{A}$ denotes a collective natural strategy. The complexity of a natural collective strategy is the sum of the individual strategies' complexities. For further insights into the semantics and theoretical aspects, refer to \cite{murano2019natural}.

\paragraph{Concurrent Game Structure.}
Formally, a \textit{CGS} consists of a tuple $S = \langle Agt, Q, \allowbreak AP, \pi, d, \delta \rangle$ with the following components:
\begin{inparaenum}[(i)]
    \item $Agt \geq 1$ is a natural number indicating the number of agents;
    \item $Q$ is a finite set of states;;
    \item $AP$ is finite set of propositions;
    \item $\pi$ is a labelling function such that for each state $q$, $\pi(q) \subseteq AP$ is a set of true propositions at $q$;
    \item $d_a(q)$ indicates for each agent $a$ and state $q$, the set of actions available in the state $q$ for $a$. A \textit{move vector} at $q$ consist of a tuple $\langle j_1,\ldots, j_{Agt} \rangle$ such that $j_a \in d_a(q)$ for each agent $a$;
    \item $\delta$ is a transition function that for each state $q$ and each move vector, returns a state that results from $q$ if each agent $a$ chooses the move in the vector.
\end{inparaenum}

\paragraph{Natural Strategies.}
Given a CGS $S$ and a set of agents, a 
\setlength\intextsep{0pt}
\begin{wrapfigure}{r}{0.2\textwidth}
    \includegraphics[width=1\linewidth]{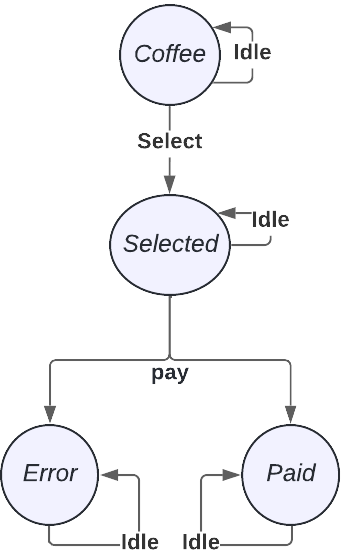}
    \caption{Vending Machine Graph}
    \label{fig:VMgraph}
\end{wrapfigure}
strategy for an agent in a \textit{MAS} is a function that determines a move for each agent based on every finite prefix of a computation. For a specific state $q$, a coalition, and a set of strategies, the \textit{outcomes} from $q$ are all possible future sequences of states from $q$ that the coalition can enforce by following their strategies.
\textit{Natural strategies} are expressed using a rule-based system,
where each rule comprises a condition and an action. A memoryless strategy (\textit{nr}-strategy) differs from a strategy with recall (\textit{nR}-strategy) in that the former defines its conditions using Boolean formulas over AP—considering current state information only. In contrast, the latter describes its conditions through regular expressions over AP, incorporating the history of states (i.e., the sequence of states that has occurred so far). The complexity of natural strategies is gauged by the overall size of the conditions' representation.

\begin{example}
Consider a coffee vending machine (Fig. \ref{fig:VMgraph}) and its \textit{nr}-strategy: (($\lnot$ coffee $\land\lnot$ selected $\land\lnot$ paid $\land\lnot$ error, \textit{select}); (selected, \textit{pay}); ($\top$, \textit{idle})). It delineates a valid plan for users to purchase the product. The complexity of this strategy, following its definition above, is $13$.
\end{example}




\paragraph{Satisfaction Relation.}
A state $q$ satisfies an \textit{NatATL} formula $\varphi$ in the structure S if there exists a sequence of states starting in $q$, that lead where the formula holds true. Formally, it is written $S,q \models \varphi$ to indicate that the state $q$ satisfies the formula $\varphi$. The satisfaction relation $\models$ is inductively defined for all states $s$ of $S$ and propositions $p \in AP$.

\section{Implementation}
Our \textit{NatATL} module enhances the functionality of the \textit{VITAMIN} tool\footnote{Visit https://vitamin-app.streamlit.app to execute and try the \textit{NatATL} module.} \cite{vitamin}. Similar to \textit{VITAMIN}, two inputs are required for this tool: a text file that represents the \textit{CGS} as an adjacency matrix - including information such as states, agents, and atomic propositions along with a labelling matrix that marks propositions' held states - and a logical formula to be verified.

\begin{figure}[h]
    \centering
    \includegraphics[width=0.97\linewidth]{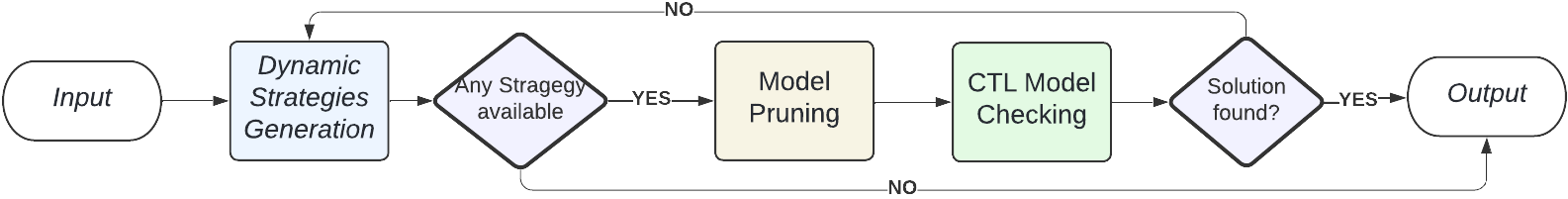}
    \caption{\textit{NatATL} Model Checking process}
    \label{fig:flowchartNatATL}
\end{figure}

Our tool consists of three main Python procedures: Strategies Generation, Model Pruning, and \textit{CTL} Model Checking. It starts by checking each possible strategy one by one. For each collective strategy $s_A$, the input model is pruned by removing all transitions that do not conform to $s_A$, transforming the system into a Kripke structure. Then, standard model checking algorithms for \textit{CTL} are applied, leveraging their polynomial complexity to achieve efficient verification. A comprehensive overview of the entire process is shown in Fig. \ref{fig:flowchartNatATL}.

\FloatBarrier

\subsection{Strategies Generation}
In this subsection, we focus on the generation of natural strategies. We present Algorithm \ref{alg1}, designed to produce optimal strategies. The generation of strategies poses a significant challenge due to the vast number of possible strategies and their varying complexities. It is essential to consider all potential strategies that the coalition under examination can produce. To tackle this challenge, we have optimized the process by implementing a dynamic strategy generation function (lines 4-18 of Algorithm \ref{alg1}). This function incrementally generates strategies, starting from those with minimal complexity (the lowest $k$ value) and only progressing to higher $k$ values if no solutions are found in simpler instances, miming a fixed approach. This dynamic, incremental approach proves optimal for two main reasons: \textit{computational efficiency} and \textit{optimization}. By focusing on lower $k$ values first, the function generates fewer strategies compared to the theoretical approach that considers \textit{always} all strategies up to a fixed upper bound. This segmentation streamlines the identification of the smallest winning strategy, reducing the time required to find an optimal solution.
The total time complexity of the generate\_strategies() algorithm is $O(n \cdot m^{k} \cdot k) + O(2^{n})$, where $n$ is the number of agents, $m$ is the number of possible actions per agent, and $k$ is the parameter for the combination length. This complexity arises due to the generation of all possible action combinations and the recursive search for collective strategies. The space complexity is similarly exponential due to the storage requirements for all potential strategies and their comparisons, primarily driven by the number of agents and actions involved\footnote{Optimizing this algorithm can be challenging due to its inherently combinatorial nature, and even if potentially heuristics, approximation, or constraint-based approaches could enhance its performance, this work aims to set the fundamentals for later adjustments.}. Despite its complexity, selecting a random collective strategy $s_A$ allows for model pruning by removing edges that conflict with $s_A$. This process is detailed in the following paragraph.

\begin{algorithm}[H]
	\renewcommand{\baselinestretch}{0.8}
	\captionsetup{font=footnotesize} 
	\caption{Dynamic Strategies Generation}\label{alg1}
	\begin{lstlisting}[language=Python, numbers=left, stepnumber=1, xleftmargin=2em]
		function generate_strategies(guarded_actions, k, agents):
		initialize an empty list of agents strategies
		
		function search(strategies):
		if all agents have strategies:
		return current strategy
		else:
		for each available strategy for the current agent:
		add it to the list of strategies
		explore further strategies recursively
		if a solution has not been found yet:
		for each agent:
		for each possible condition-action rule:
		if condition complexity less or equal than k:
		create a new strategy
		if it is not already in the list:
		add it to the agents list of strategies
		continue searching recursively
		
		return strategies
	\end{lstlisting}
\end{algorithm}

\subsection{Model Pruning}
This task involves projecting each collective strategy generated by Algorithm \ref{alg1} onto the model. The model is pruned based on the actions chosen by the agents in the coalition $A$, removing all moves incompatible with the examined strategies. To clarify this phase, we distinguish between two approaches: model pruning for memoryless strategies and model pruning for strategies with recall.

\paragraph{nr-Strategies Model Pruning.}
For each selected $s_A$, we project the strategies only onto the model states where the current strategy conditions are met. To ensure that each agent always has an action to execute, even when none of the conditions in its current strategy are satisfied, we include an additional condition-action pair, $(\top, \textit{idle})$. This pair allows the agent to perform the idle action by default in any state, correlating to its true condition (i.e., applicable to all states) whenever no other action can be performed. If a current strategy includes an action that is not present in the current model states (where its conditions are met), the collective strategy is discarded as invalid.
\vspace{1em}
\begin{algorithm}[H]
\renewcommand{\baselinestretch}{0.8}
\captionsetup{font=footnotesize} 
\caption{Model Pruning - Memoryless Approach}\label{alg2}
\scriptsize
\begin{lstlisting}[language=Python, numbers=left, stepnumber=1, xleftmargin=2em]
function prune_model(CGS, agents):
 create a copy of the CGS
 initialize a list to track modified rows
  for each row in the copied CGS:
   initialize a flag to track row pruning
   for each element in the row:
    if the condition matches specified states:
     if the element is not empty:
      filter elements based on agent action
      update the copied CGS
      mark the row as modified
  return the modified CGS
\end{lstlisting}
\end{algorithm}
\vspace{1em}
The implemented algorithm is outlined in Algorithm \ref{alg2}. Our algorithm begins by duplicating the original model, interpreted as a transition matrix, to preserve the initial data (line 1). It then systematically identifies and records the rows subject to pruning (lines 4-5). Each row and its individual elements, representing transitions, are meticulously inspected. A comparison is made between the action prescribed for the coalition and those specified within the element. Elements with matching actions are retained, while those diverging from the agents' designated action are eliminated (line 9). Upon completing this iterative process, the algorithm returns the pruned model. Any checks for the validity of the strategy are performed by external functions.
The total time and space complexity of the modify\_matrix() algorithm results to be $O(n^2)$ where $n$ is the number of nodes in the graph. This complexity arises from the need to iterate over and potentially modify every pair of nodes in the graph, leading to quadratic scaling with respect to the number of nodes.

\paragraph{nR-Strategies Model Pruning.}
Unlike the theoretical approach that involves the use of Automata and a different method for solving model checking for \textit{NatATL} with recall (see \cite{natatl2019} for more details), in this work, we propose a different approach that involves using tree structures instead of automata. This approach does not undermine the method shown with memoryless strategies but rather leverages the potential of this representation to capture the concept of history. 
\vspace{1em}
\begin{algorithm}[H]
\renewcommand{\baselinestretch}{0.8}
\captionsetup{font=footnotesize} 
\caption{CGS into Tree Conversion Function}\label{alg3}
\begin{lstlisting}[language=Python, numbers=left, stepnumber=1, xleftmargin=2em]
function build_tree(CGS, model_states, height):

 function add_children(node, current_level):
  if maximum depth (height) reached:
   return
  get the current state's index from the node
  get the possible transitions for this state from CGS
  for each possible next state and actions:
   if there are no actions: 
    skip to the next
   get the next state using its index
   for each action:
   if the next state isn't already in our nodes:
    create it
   create a new child node for the next state and action
   add this child to the current node.
   add the new node to our list of nodes.
   add_children(new child, current_level + 1)

 create the root node with model initial state
 start the list of nodes with the root node
 add_children(root, 1)
 return root node
    
\end{lstlisting}
\end{algorithm}
\vspace{1em}
From a practical and algorithmic standpoint, using a tree can be preferred over an automaton for several reasons. Firstly, tree structures can naturally represent hierarchical data and branching processes, which aligns well with the concept of history and state transitions in many applications. Secondly, trees can provide more efficient traversal and manipulation algorithms, allowing for quicker updates and searches. Lastly, trees facilitate a more intuitive visualization and understanding of the state space, making it easier to debug and optimize the model. 
So, for the model pruning approach in this context, we unrolled the initial \textit{CGS} into a tree structure to better represent the concept of histories and ensure that we keep track of all transitions and the different paths that arise from the base model, thus taking repetitions into account. We present the conversion function created specifically for the model extracted from the initial input file in Algorithm \ref{alg3}.
Our algorithm constructs a tree from a \textit{CGS} where nodes represent states and edges represent transitions. It starts by defining a helper function (line 3) to recursively add child nodes. If the current level reaches the specified height (line 4), the function stops, limiting the tree's depth. The function finds the source state and transitions from the \textit{CGS} (line 7), identifying the next state and actions. If the next state is not already in the tree, a new node is created (lines 13-14). The process begins by creating the root node for the initial state and calling add\_children on the root node to build the tree (lines 20-21). Once all nodes are added up to the specified height, the root node is returned (line 23), representing the complete tree with all possible transitions and actions up to the given depth. The algorithm has a time complexity of \(O(n^{\text{height}} \cdot a)\) and space complexity of \(O(n^{\text{height}})\), where \(n\) is the number of states and \(a\) is the average number of actions per transition. Each node tracks the state, the action leading to it, its children, its history, propositions' truth values, and a pruning flag. After model conversion, we project the current collective strategy onto the tree. Each individual strategy within a collective strategy must be validated before pruning. Strategies with actions not present in the model for states where the condition is met are considered invalid. Starting from a collective strategy, each individual strategy prunes the tree, potentially reducing its size for the next iteration and the next agent's strategy. A valid collective strategy may become invalid for a reduced model, requiring a correctness check each time an agent operates on an individual strategy. Pruning depends on the current guarded action's condition, whether a boolean formula or a regular expression. For regular expressions, pruning occurs along the nodes satisfying the expression's path. We devised an algorithm for this (see Algorithm \ref{regexalgo}).
\vspace{0.5em}
\begin{algorithm}[H]
\renewcommand{\baselinestretch}{0.8}
\captionsetup{font=footnotesize} 
\caption{Regex Pruning Algorithm}\label{regexalgo}
\begin{lstlisting}[language=Python, numbers=left, stepnumber=1, xleftmargin=2em]
function regex_path(tree, height, model, strategies):
 traverse tree to find regex-matching path up to height
 if no path matches the regex condition: return False
 function prune_nodes_along_path(node, path, action, current_level):
  initialize a flag entered_statement as False
  if current level < path length and node state matches path state:
   initialize an empty list to_remove
   for each child and corresponding action of the node:
    if strategy action != specified action and node not pruned:
     add the index of the child to the list to_remove
     set entered_statement to True
   if entered_statement is True:
    mark the node as pruned
   for each index in the list to_remove in reverse order:
    remove the corresponding child and action from the node
   for each remaining child of the node:
    if pruning the child returns True:
     set entered_statement to True
  return entered_statement
  call prune_nodes_along_path from root and return its result

\end{lstlisting}
\end{algorithm}
Algorithm \ref{regexalgo} finds a path in the tree that matches a given regular expression condition (lines 2-3) and prunes nodes along that path that do not meet the specified action for a given strategy. Pruning involves identifying and removing child nodes that do not match the action (lines 8-15), applied recursively to all tree levels (line 20). 
\vspace{0.5em}
\begin{algorithm}[H]
\renewcommand{\baselinestretch}{0.8}
\captionsetup{font=footnotesize} 
\caption{Model Pruning Algorithm}\label{alg4}
\begin{lstlisting}[language=Python, numbers=left, stepnumber=1, xleftmargin=2em]
function pruning(tree, height, model, strategies):
 initialize flag as False
 for each collective strategy in strategies:
  start with the first individual strategy
  for each guarded action in the agent s strategy:
   check whether the current guarded action strategy is valid
   if strategy is valid:
    determine whether the condition is a regex or a boolean formula
    if it is a boolean formula:
     perform boolean pruning on the tree
     if not all nodes are pruned:
      set the flag to False
   else
    return False due to invalid strategy
  if no agent actions available:
   perform idle pruning on the tree
  reset the flags used for pruning
 rename nodes in the tree
 transform the pruned tree into an extended CGS
 update the new CGS file with the modified datae
 perform CTL model checking on the new model
    
\end{lstlisting}
\end{algorithm}
\vspace{0.5em}
If the condition is a boolean formula, all subtrees from paths inconsistent with the selected action are removed for nodes whose state matches the condition. A flag is set to True for each node leading to pruning, preventing repeated pruning within the same individual strategy. When switching to a new strategy for a different agent, the pruning flags are reset to False. To ensure an agent can act each turn, if no guarded action is applicable within a strategy, a supplementary condition-action pair ($\top$, idle) is added, allowing each agent to perform idle in every state. This ensures at least one possible action per turn. The initial \textit{CGS} construction ensures each agent can perform idle in every state. The pruned tree is reconverted into an extended \textit{CGS}, with unique state renaming to avoid repetition, resulting in the desired \textit{CGS} for \textit{CTL} model checking. The main pruning algorithm summarizing this process is presented in Algorithm \ref{alg4}. The time complexity of the pruning algorithm is \( O(n \cdot m \cdot (V + E + M)) \), where \( n \) represents the number of strategies, \( m \) denotes the number of conditions per strategy, \( V \) is the number of nodes in the tree, \( E \) is the number of edges in the tree, and \( M \) is the number of states in the model. The space complexity of the algorithm is \( O(V + E + n \cdot m + M) \), accounting for the storage of the tree, the strategies, and the model states.

\subsection{CTL Model Checking}

The \textit{CTL} model checking algorithm will therefore be executed by taking as input a \textit{CTL} formula (properly derived through a specific function that will transform the initial \textit{NatATL} formula into its equivalent universally quantified), and the new model obtained from the previous pruning step. The \textit{CTL} model checking algorithm implemented in this code is based on the theoretical foundations presented in \cite{jamroga2015logical}, which can be consulted for further details. 
\setlength\intextsep{0pt}
\begin{wrapfigure}{r}{0.2\textwidth}
    \includegraphics[width=0.9\linewidth]{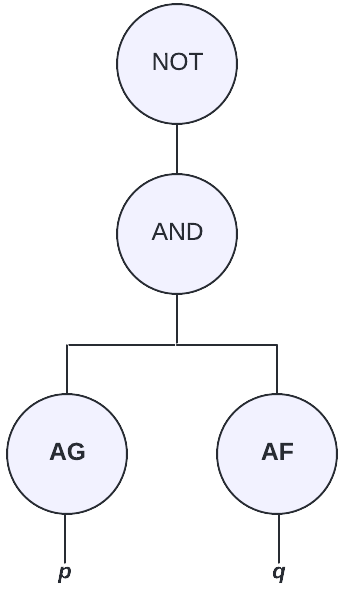}
    \caption{Formula tree $\neg((AG \ p) \ \land \ (AF \ q))$}
    \label{fig:tree}
\end{wrapfigure}
Nevertheless, to provide some basic foundations, the standard model checking algorithm for \textit{CTL} verifies temporal properties using fixpoint computation. It translates logical operations into set-theoretic operations and modal operators into their corresponding pre-images. Long-term properties are checked through fixpoint computation. The \textit{Next} operator is crucial, as it applies to existential and universal strategic operators (EX, AX) for long-term modalities (EU, AU, EG, AG). Our algorithm employs a tree structure to store the parsed formula, using both a depth-first search approach to reach leaves and a bottom-up technique to handle even logic formulas with more temporal operators. Each node represents a sub-formula, eventually satisfied by a set of states that the algorithm calculates. If a solution validating the formula is found, the algorithm returns the set of states satisfying the root of the formula tree\footnote{We consider a formula tree as a tree structure that stores a parsed logical formula.} (refer to Fig. \ref{fig:tree} for a visual representation of a formula tree) along with the winning strategy. In case the ongoing model verification fails to yield a solution for the pruned model, the algorithm regresses to its initial Strategies Generation phase. It then proceeds to project subsequent strategies onto the original model and reiterates the model checking process. This iterative cycle continues until either a solution is found or no further strategies remain. The algorithms reported below (Algorithm \ref{modelChecking1}, \ref{modelChecking2}) encapsulate respectively the main function of model checking and the core of the whole model checking process. In fact, the former handles the model checking functions and results, while the latter executes the formal verification of the formula upon the model (due to space constraints, we have chosen to include only the representation of the algorithm with a single path quantifier. For the complete representation, please visit the project directory at the following \href{https://anonymous.4open.science/r/NatATL-Tool-PRIMA-2024-7CD7}{\textit{link}}).
\vspace{0.5em}
\begin{algorithm}[H]
\renewcommand{\baselinestretch}{0.8}
\captionsetup{font=footnotesize} 
\caption{Model Checking}
\label{modelChecking1}
\scriptsize
\begin{lstlisting}[language=Python,numbers=left, stepnumber=1, xleftmargin=2em]
function model_checking(formula, model):
 parse model and formula
 build the formula-tree
 if tree root is empty:
  return error
 perform CTL model checking
 retrieve initial state
 verify whether it satisfies the formula against tree root
 return result

\end{lstlisting}
\end{algorithm}
\begin{algorithm}[H]
\renewcommand{\baselinestretch}{0.8}
\captionsetup{font=footnotesize} 
\caption{CTL Formula Verification upon the Model}
\label{modelChecking2}
\scriptsize
\begin{lstlisting}[language=Python,numbers=left, stepnumber=1, xleftmargin=2em]
function solve_tree(node, formula):
 if Formula is a proposition:
  return set of nodes where the proposition holds
 if Formula is of the form NOT Subformula:
  Subresult = solve_tree(Tree, Subformula)
  return complement of Subresult in the set of all nodes
 if Formula is of the form Subformula1 AND Subformula2:
  Result1 = solve_tree(Tree, Subformula1)
  Result2 = solve_tree(Tree, Subformula2)
  return intersection of Result1 and Result2

 if Formula is of the form AX Subformula:
  Subresult = solve_tree(Tree, Subformula)
  return universal_preimage(Subresult)

 if Formula is of the form AG Subformula:
  Q1 = set of all nodes
  Q2 = solve_tree(Tree, Subformula)        
  while Q1 is not a subset of Q2:
   Q1 = Q2
   Q2 = intersection of Q1 and universal_preimage(Q1)
  return Q1

 if Formula is of the form A(Subformula1 U Subformula2):
  Q1 = empty set
  Q2 = solve_tree(Tree, Subformula1)
  Q3 = solve_tree(Tree, Subformula2)        
  while Q3 is not a subset of Q1:
   Q1 = union of Q1 and Q3
   Q3 = intersection of Q2 and universal_preimage(Q1)   
  return Q1
    
 return empty set

\end{lstlisting}
\end{algorithm}
\vspace{0.5em}
The time complexity of the \textit{CTL} model checking function solve\_tree is $O(|\varphi| \cdot |S|^2)$, where $|\varphi|$ is the size of the formula and $|S|$ is the number of states in the model. This is because the function recursively traverses the formula tree ($O(|\varphi|)$) and performs operations such as finding predecessors for each state, which can take up to $O(|S|^2)$ time in the worst case. These operations are needed for each node in the formula tree, leading to the overall time complexity. The space complexity of the same function is $O(|\varphi| \cdot |S|) + O(|S|^2)$. The space is needed for storing the formula tree, which requires ($O(|\varphi|)$) space, and for storing the sets of states at each node, which can take up to $O(|\varphi| \cdot |S|)$ space. Additionally, storing the state transitions and predecessor sets requires $O(|S|^2)$ space. Therefore, the total space complexity is influenced by both the formula size and the number of states in the model. These complexities are considered optimal for the problem of model checking using \textit{CTL}. The linear dependence on the formula size is the best achievable, as each part of the formula must be evaluated. The quadratic dependence on the number of states arises from the necessity to explore all state transitions, which inherently involves checking relationships between pairs of states.

\section{Experiments}
We have performed extensive tests for the proposed verification tool on an HP Omen 15-ax213ng with an Intel i7-7700HQ 3.8 GHz CPU and 16 GB of RAM. The software development for this project is conducted using Python 3.9 and the PyCharm IDE. The software hosting service is provided by GitHub, facilitating the integration of this work into the existing \textit{VITAMIN} checker.

Regarding the tests conducted on our tool, an initial series was carried out to verify the accuracy of the responses provided by the \textit{NatATL} model checking algorithm. We used the \textit{ATL} model checking module of \textit{VITAMIN} as a benchmark since this work introduces the first implementation of a \textit{NatATL} verification module, and a real benchmark for performance comparison does not yet exist. To initiate a preliminary experiment, it is crucial to examine the correspondence between the Boolean solutions generated by the two algorithms. Ideally, there should be complete equivalence between their results. However, in practice, \textit{NatATL}'s behavior can vary depending on the input model and/or formula due to a complexity bound on the generated strategies. Specifically, if \textit{ATL} produces a \textit{True} result, \textit{NatATL} may yield a \textit{False} result up to a certain $k$ value, as it provides the smallest winning strategy. It is necessary to determine this minimum complexity bound for the strategy. Each temporal operator was tested a hundred times to understand its behavior. Identical inputs were used for both \textit{NatATL} and \textit{ATL}, with their corresponding formula versions. For computational purposes, the \textit{NatATL} complexity bound was increased up to 10, considering it challenging for a human to track more than 10 guarded actions for a single strategy. In a comprehensive analysis of nearly a thousand tests, it was conclusively demonstrated that the \textit{NatATL} verification system operates correctly. This is because \textit{NatATL}, as an \textit{ATL} update, will produce the same result for the same model. Therefore, this test is crucial in certifying the soundness of the implemented algorithm.

\begin{figure}[h]
    \centering
    \includegraphics[width=0.97\linewidth]{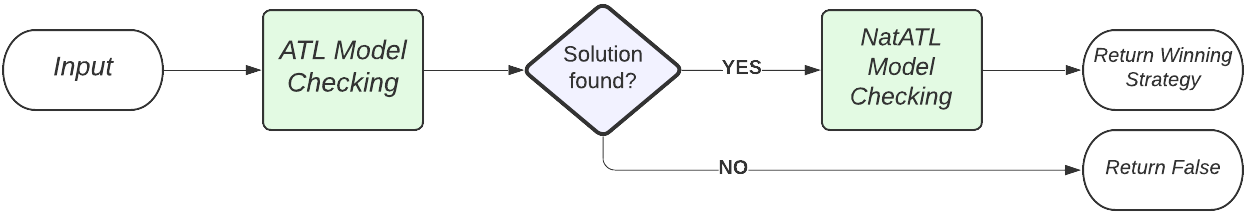}
    \caption{\textit{NatATL} Model Checking process}
    \label{fig:flowchartNatATL2}
\end{figure}

To assess the time performance of our tool, we conducted a thousand more tests, systematically varying the number of states (i.e., model size), the complexity bound, and the number of agents within the formula coalition. Our analysis revealed that the execution behavior is significantly influenced by the number of agents and the complexity bound. Each agent requires a specific set of strategies based on their potential actions, so increasing these values leads to longer execution times. Nevertheless, the tool demonstrates notable scalability regarding the number of states, maintaining consistent response times even for models with millions of states. Graphical results for the memoryless strategies can be seen in Fig. \ref{table1}, \ref{table2.1}.
For tests conducted on the tool with recall, we set the maximum height of the structure to a predefined value of 5 for computational efficiency \footnote{Beyond a depth of 5, the tool started to experience slight delays in generating the entire tree, so we chose this height to maintain real-time processing.}. Thus, we observed slowdowns due to the necessity of representing histories. The execution time was found to be directly proportional to the degree of the tree, significantly influenced by the density of the transition matrix, particularly the number of different transitions from the same source state to the same destination state. Higher density led to longer execution times. Conversely, sparser matrices with lower tree degrees resulted in shorter execution times for this case. 
\vspace{1em}
\begin{figure}[h]
\centering
\begin{minipage}[b]{0.45\linewidth}
\centering
\begin{tikzpicture}
\begin{axis}[
xlabel={Number of States},
ylabel={Average Time (seconds)},
xmode=log,
log basis x=10,
xmin=1,
xmax=10000,
ymin=0, ymax=1,
ytick={0, 0.5, 1},
xtick={1, 10, 100, 1000, 10000},
xticklabels={$10^0$, $10^1$, $10^2$, $10^3$, $10^4$},
xticklabel style={font=\tiny, /pgf/number format/1000 sep={}},
yticklabel style={font=\tiny},
legend pos=outer north east,
scale=0.35,
width=350pt, height=250pt,
bar width=7pt,
]
\addplot[fill=cyan!40, draw=cyan!60!black, ybar] coordinates {
(10,0.03)
(100,0.05)
(1000,0.10)
(10000,0.11)
(100000,0.14)
};
\end{axis}
\end{tikzpicture}
\caption{Average response times for number of states and $|A| = 1$}
\label{table1}
\end{minipage}
\hfill
\begin{minipage}[b]{0.45\linewidth}
\centering
\begin{tikzpicture}
\begin{axis}[
xlabel={Number of States},
ylabel={},
xmode=log,
log basis x=10,
xmin=1,
xmax=10000,
ymin=0, ymax=1,
ytick={0, 0.5, 1},
xtick={1, 10, 100, 1000, 10000},
xticklabels={$10^0$, $10^1$, $10^2$, $10^3$, $10^4$},
xticklabel style={font=\tiny, /pgf/number format/1000 sep={}},
yticklabel style={font=\tiny},
legend pos=outer north east,
scale=0.35,
width=350pt, height=250pt,
bar width=7pt,
]
\addplot[fill=cyan!40, draw=cyan!60!black, ybar] coordinates {
(10,0.15)
(100,0.17)
(1000,0.19)
(10000,0.21)
(100000,0.24)
};
\end{axis}
\end{tikzpicture}
\caption{Average response times for number of states and $|A| = 4$}
\label{table2.1}
\end{minipage}
\end{figure}
\vspace{0.5em}
More specifically, we adopted two verification approaches: the first aimed at verifying model robustness, considering nondeterministic models where the matrix density was maximal and multiple occurrences of the same elements on the same row were present (hence nondeterminism). The second approach focused solely on strategy robustness, adopting classical incomplete models in multi-agent scenarios and discarding potentially invalid strategies for that non-robust model. The first approach empirically set an upper bound on the temporal complexity due to the analysis of input models that consider all possible combinations of actions characterizing different transitions for each row (and for each element within it), addressing the worst-case scenario, which is less likely in practice. The second approach reflects the average case where the user generates an incomplete model due to scenarios in multi-agent systems, inevitably leading to lower and more efficient execution times compared to the worst-case scenario. The tables in the Fig. \ref{table3.1}, \ref{table3} summarize the time analysis conducted.
\vspace{0.5em}
\begin{figure}[h]
\centering
\begin{minipage}[b]{0.45\linewidth}
\centering
\begin{tikzpicture}
\begin{axis}[
    ylabel style={font=\scriptsize},
    xlabel={Number of States},
    ylabel={Average Time (seconds)},
    xlabel style={font=\small},
    xmode=log,
    log basis x=10,
    xmin=1, xmax=1000,
    ymin=0, ymax=7200,
    ytick={0, 1000, 2000, 3000, 4000, 5000, 6000, 7000}, 
    xtick style={font=\tiny},
    ytick style={font=\tiny},
    tick label style={font=\tiny},
    scale=0.35,
    width=350pt, height=250pt,
    legend style={font=\tiny, fill=none, at={(0.05,0.95)}, anchor=north west},
    ]
\addplot[blue,dashed,mark=square*,mark options={solid},line width=1.5pt] coordinates {
    (1,0)
    (10,10.58) 
    (100,180) 
    (1000,3600) 
};
\addplot[red,solid,mark=*,line width=1.5pt] coordinates {
    (1,0)
    (10,20.12) 
    (100,360) 
    (1000,7200)
};
\legend{Sparse, Dense}
\end{axis}
\end{tikzpicture}
\caption{Average case: Average time table comparison using dense and sparse transition matrix and $|A|=3$}
\label{table3.1}
\end{minipage}
\hfill
\begin{minipage}[b]{0.45\linewidth}
\centering
\begin{tikzpicture}
\begin{axis}[
    ylabel style={font=\scriptsize},
    xlabel={Number of States},
    ylabel={Average Time (seconds)},
    xlabel style={font=\small},
    xmode=log,
    log basis x=10,
    xmin=1, xmax=1000,
    ymin=0, ymax=7200,
    ytick={0, 1000, 2000, 3000, 4000, 5000, 6000, 7000}, 
    xtick style={font=\tiny},
    ytick style={font=\tiny},
    tick label style={font=\tiny},
    scale=0.35,
    width=350pt, height=250pt,
    legend style={font=\tiny, fill=none, at={(0.05,0.95)}, anchor=north west},
    ]
\addplot[blue,dashed,mark=square*,mark options={solid},line width=1.5pt] coordinates {
    (1,0)
    (10,3600) 
    (100,7200) 
    (1000,7200)
};
\addplot[red,solid,mark=*,line width=1.5pt] coordinates {
    (1,0)
    (10,5400) 
    (100,7200) 
    (1000,7200)
};
\legend{Sparse, Dense}
\end{axis}
\end{tikzpicture}
\caption{Worst case: Average time table comparison using dense and sparse transition matrix and $|A|=3$}
\label{table3}
\end{minipage}
\end{figure}
\paragraph{Further Analysis.}
Despite its potential, the tool's real-world applicability is limited by instances where model checking fails. If a formula is not satisfied by the model, the tool must generate all strategies up to a fixed complexity bound before confirming the negative result, leading to longer response times compared to positive instances. Our strategy exploration approach, which prioritizes generating strategies with lower complexity and gradually increasing to the fixed bound, reduces execution time. Testing revealed that solutions, if they exist, are typically found at lower complexity levels, even with a fixed bound of 10 (see Fig. \ref{table2}, \ref{table2.2}).
\vspace{0.5em}
\begin{figure}[h]
\vspace{-0.5em}
\centering
\begin{minipage}[b]{0.45\linewidth}
\centering
\begin{tikzpicture}
\begin{axis}[
    xlabel={Complexity Bound},
    ylabel={Average Time (seconds)},
    xlabel style={font=\small},
    ylabel style={font=\small},
    xmin=0, xmax=10,
    ymin=0, ymax=1,
    xtick={0, 1, 2, 3,4,5,6,7,8,9, 10},
    xticklabels={0, 1, 2,3,4,5,6,7,8,9, 10},
    ytick={0,0.5,1}, 
    xticklabel style={font=\tiny, /pgf/number format/1000 sep={}},
    yticklabel style={font=\tiny},
    legend pos=outer north east,
    scale=0.35,
    width=350pt, height=250pt,
    legend style={font=\footnotesize}
    ]
\addplot[blue, mark=*, only marks] coordinates {
    (1,0.015) (1,0.025) (1,0.012) (1,0.018) (1,0.011) (1,0.013) (1,0.025) (1,0.031)
    (1,0.020) (1,0.033) (1,0.013) (1,0.010) (1,0.031) (1,0.021) (1,0.010) (1,0.032)
    (1,0.012) (1,0.020) (1,0.022) (1,0.013) (1,0.019) (1,0.010) (1,0.030) (1,0.034)
    (1,0.018) (1,0.027) (1,0.011) (1,0.038) (1,0.021) (1,0.011) (1,0.021) (1,0.021)
    (1,0.035) (1,0.022) (1,0.011) (1,0.028) (1,0.014) (1,0.018) (1,0.010) (1,0.033)
    (1,0.034) (1,0.023) (1,0.011) (1,0.037) (1,0.026) (1,0.031) (1,0.022) (1,0.021)
    (1,0.018) (1,0.011) (1,0.032) (1,0.011) (1,0.016) (1,0.015) (1,0.023) (1,0.021)
    };
    
\addplot[blue, mark=*, only marks] coordinates {
    
    (2,0.334)
    (2,0.366)
    (2,0.338)
    (2,0.362)
    (2,0.331)
    (2,0.367)
    (2,0.365)
    (2,0.36)
    (2,0.335)
    (2,0.368)
    (2,0.369) 
    (2,0.362)
    (2,0.331)
    (2,0.35)
    (2,0.33)
    (2,0.345) 
    (2,0.355) 
    (2,0.36)
    (2,0.331)
    (2,0.361)
    (2,0.341)
    (2,0.362)
    (2,0.342)
    (2,0.335)
    (2,0.351)
    (2,0.36)
    };
\addplot[blue, mark=*, only marks] coordinates {
    (3,0.741) (3,0.759) (3,0.705)
    };
\end{axis}
\end{tikzpicture}
\captionof{figure}{Time-to-solution table with $k = 10$ and $|A| = 1$}
\label{table2}
\end{minipage}
\hfill
\begin{minipage}[b]{0.45\linewidth}
\centering
\begin{tikzpicture}
\begin{axis}[
    xlabel={Complexity Bound},
    ylabel={},
    xlabel style={font=\small},
    ylabel style={font=\small},
    xmin=0, xmax=10,
    ymin=0, ymax=3000,
    xtick={0, 1, 2, 3,4,5,6,7,8,9, 10},
    xticklabels={0, 1, 2,3,4,5,6,7,8,9, 10},
    ytick={0,1500,3000}, 
    xticklabel style={font=\tiny, /pgf/number format/1000 sep={}},
    yticklabel style={font=\tiny},
    legend pos=outer north east,
    scale=0.35,
    width=350pt, height=250pt,
    legend style={font=\footnotesize}
    ]
\addplot[blue, mark=*, only marks] coordinates {
    (1,0.115) (1,0.225) (1,0.152) (1,0.178) (1,0.141) (1,0.113) (1,0.145) (1,0.131)
    (1,0.150) (1,0.133) (1,0.153) (1,0.160) (1,0.149) (1,0.141) (1,0.152) (1,0.132)
    (1,0.162) (1,0.166) (1,0.122) (1,0.113) (1,0.119) (1,0.157) (1,0.170) (1,0.134)
    (1,0.118) (1,0.227) (1,0.171) (1,0.178) (1,0.151) (1,0.181) (1,0.121) (1,0.121)
    (1,0.135) (1,0.122) (1,0.165) (1,0.128) (1,0.114) (1,0.218) (1,0.158) (1,0.133)
    (1,0.134) (1,0.123) (1,0.111) (1,0.167) (1,0.126) (1,0.131) (1,0.162) (1,0.181)
    (1,0.148) (1,0.161) (1,0.152) (1,0.161) (1,0.136) (1,0.145) (1,0.153) (1,0.163)
    };
    
\addplot[blue, mark=*, only marks] coordinates {
    (2,1132.334)
    (2,1131.446)
    (2,1129.338)
    (2,1133.362)
    (2,1128.331)
    (2,1135.367)
    (2,1132.365)
    (2,1134.36)
    (2,1128.335)
    (2,1136.368)
    (2,1127.369) 
    (2,1130.362)
    (2,1132.331)
    (2,1131.35)
    (2,1126.33)
    (2,1138.345) 
    (2,1133.355) 
    (2,1132.36)
    (2,1131.331)
    (2,1134.361)
    (2,1130.341)
    (2,1131.362)
    (2,1132.342)
    (2,1133.335)
    (2,1132.351)
    (2,1132.36)
    };
\addplot[blue, mark=*, only marks] coordinates {
    (3,2962.831) (3,2959.777) (3,2960.100)
    };
\end{axis}
\end{tikzpicture}
\captionof{figure}{Time-to-solution table with $k = 10$ and $|A| = 4$}
\label{table2.2}
\end{minipage}
\end{figure}
\vspace{0.5em} 
Furthermore, by intelligently managing strategy complexity and avoiding negative instances, we improve the tool's usability. Running the algorithm only for positive instances, where the formula is satisfied, allows for faster responses, yielding the optimal minimum-complexity winning strategy by exploiting its ability to synthesize strategies. We can achieve this by initially validating instances with a preliminary \textit{ATL} model checking. The flowchart in Figure \ref{fig:flowchartNatATL2} demonstrates the tool's practical application: it starts by taking a model and an \textit{ATL} formula as input, to subsequently perform \textit{ATL} model checking, terminating if the instance is unsatisfiable, otherwise, translating the \textit{ATL} formula into its \textit{NatATL} corresponding to execute the \textit{NatATL} model checking procedure, which outputs the optimal winning strategy. 

\vspace{1em}
\FloatBarrier

\vspace{1em}
To evaluate our approach, Fig. \ref{table4}, \ref{table4.1} present a comparison of average response times between traditional \textit{NatATL} utilization and the combined $\allowbreak ATL+NatATL$ method both for memoryless and with recall approaches. We tested unsatisfiable formulas with a fixed complexity bound and three agents involved. The practical viability of our tool is illustrated via blue line, representing the optimal response times for the \textit{ATL}+\textit{NatATL} approach, compared to the red line, showing standalone \textit{NatATL} usage. As expected, our approach significantly improves response times in real-world scenarios. Negative instances highlight the time differences, requiring \textit{NatATL} to generate all possible strategies. Our new method filters real-time results without invoking the \textit{NatATL} strategy generation algorithm, demonstrating the benefits of our optimization. Time constraints naturally increase with the number of agents and complexity bound, as the computational effort to generate individual strategies for each agent remains substantial.
Specifically, for a single agent, $k$ should not exceed $10$ to maintain acceptable performance. 
\vspace{0.5em}
\begin{figure}[h]
\centering
\begin{minipage}[b]{0.45\linewidth}
\centering
\begin{tikzpicture}
\begin{axis}[
    ylabel style={font=\scriptsize},
    xlabel={Number of States},
    ylabel={Average Time (seconds)},
    xlabel style={font=\small},
    xmode=log,
    log basis x=10,
    xmin=1, xmax=1000,
    ymin=0, ymax=4000,
    ytick={0, 1000, 2000, 3000, 4000}, 
    xtick style={font=\tiny},
    ytick style={font=\tiny},
    tick label style={font=\tiny},
    scale=0.35,
    width=350pt, height=250pt,
    legend style={font=\tiny, fill=none, at={(0.05,0.95)}, anchor=north west},
    ]
\addplot[blue,dashed,mark=square*,mark options={solid},line width=1.5pt] coordinates {
    (1,0)
    (10,0.25) 
    (100,0.24) 
    (1000,0.33) 
};
\addplot[red,solid,mark=*,line width=1.5pt] coordinates {
    (1,0)
    (10,3531) 
    (100,3559) 
    (1000,3643)
};
\legend{ATL+NatATL, NatATL}
\end{axis}
\end{tikzpicture}
\caption{$\allowbreak ATL+NatATL$ times table: Memoryless approach}
\label{table4}
\end{minipage}
\hfill
\begin{minipage}[b]{0.45\linewidth}
\centering
\begin{tikzpicture}
\begin{axis}[
    ylabel style={font=\scriptsize},
    xlabel={Number of States},
    ylabel={Average Time (seconds)},
    xlabel style={font=\small},
    xmode=log,
    log basis x=10,
    xmin=1, xmax=1000,
    ymin=0, ymax=8000,
    ytick={0, 2000, 4000, 8000}, 
    xtick style={font=\tiny},
    ytick style={font=\tiny},
    tick label style={font=\tiny},
    scale=0.35,
    width=350pt, height=250pt,
    legend style={font=\tiny, fill=none, at={(0.05,0.95)}, anchor=north west},
    ]
\addplot[blue,dashed,mark=square*,mark options={solid},line width=1.5pt] coordinates {
    (1,0)
    (10,0.25) 
    (100,0.24) 
    (1000,0.33) 
};
\addplot[red,solid,mark=*,line width=1.5pt] coordinates {
    (1,0)
    (10,4531) 
    (100,7559) 
    (1000,8643)
};
\legend{ATL+NatATL, NatATL}
\end{axis}
\end{tikzpicture}
\caption{$\allowbreak ATL+NatATL$ times table: Recall approach}
\label{table4.1}
\end{minipage}
\end{figure}
\vspace{0.5em}
As the number of agents increases, the complexity bound limit needs to be reduced, indicating an implementation bottleneck due to the inherent logic. In conclusion, our combined \textit{ATL}$+$\textit{NatATL} approach provides significant performance improvements, though careful management of complexity bounds is required as the number of agents increases. This demonstrates the practical effectiveness and limitations of our method in real-world applications.

\section{Conclusions}

In this paper, we present for the first time a model checker for \textit{NatATL} along with a method for generating natural strategies, both in the case of memoryless strategies and with strategies recall. Our tool enhances strategic reasoning in \textit{MAS} by effectively addressing the limitations of traditional \textit{ATL} in capturing human decision-making processes. Our combined $ATL+NatATL$ approach demonstrates, by means of practical examples, applicability and efficiency, which makes us highly confident that our approach can be applied usefully in real-world scenarios (the latter will be a topic for future intensive works). 

Our tool ability to reduce computational effort compared to a traditional \textit{NatATL} approach is especially notable in scenarios involving strategies with recall. The results reported along the paper highlighted importance of managing complexity bounds to maintain performance as the number of agents increases. For single-agent scenarios, the complexity bound $k$ is effective up to $10$, and for multi-agent settings, careful management ensures continued efficiency. Moreover, the implementation of dynamic strategy generation and model pruning techniques contributed to the efficiency of the verification process. The segmentation of strategy generation, starting from minimal complexity and progressing only when necessary, proved optimal for both computational efficiency and resource optimization.

Despite the significant advancements, several areas for future research and development remain, such as scalability enhancements. Given the intrinsic limitations of the logic and natural strategies, scalability remains a challenging task. Additionally, the use of optimization techniques such as machine learning and heuristic-based approaches could be explored. This work provides a valuable foundation for any future tools that require the use of natural strategies, making it a significant step forward in the field of strategic reasoning in multi-agent systems. By addressing these future research areas, it will be possible to increase the tool's robustness and versatility for verifying \textit{MAS} across various domains.


%
%
%
\bibliographystyle{splncs04}
\bibliography{bibliography}

\end{document}